# A Commons-Compatible Implementation of the Sharing Economy: Blockchain-Based Open Source Mediation

Petra Unterberger[1], Manfred Mayr[2], Maximilian Tschuchnig[2] and Peter Haber[3]
[1] *Business Development & Economics, Salzburg University of Applied Sciences, Puch/Salzburg, Austria*
[2] *Business Informatics, Salzburg University of Applied Sciences, Puch/Salzburg, Austria*
[3] *Information Technology & Systems Management, Salzburg University of Applied Sciences, Puch/Salzburg, Austria*
petra.unterberger@fh-salzburg.ac.at

Abstract: The network economical sharing economy, with direct exchange as a core characteristic, is implemented both, on a commons and platform economical basis. This is due to a gain in importance of trust, collaborative consumption and democratic management as well as technological progress, in the form of near zero marginal costs, open source contributions and digital transformation. Concurrent to these commons-based drivers, the grey area between commerce and private exchange is used to exploit work, safety and tax regulations by central platform economists. Instead of central intermediators, the blockchain technology makes decentralized consensus finding, using Proof-of-Work (PoW) within a self-sustaining Peer-to-Peer network, possible. Therefore, a blockchain-based open source mediation seems to offer a commons-compatible implementation of the sharing economy. This thesis is investigated through a qualitative case study of Sardex and Interlace with their blockchain application, based on expert interviews and a structured content analysis. To detect the most commons-compatible implementation, the different implementation options through conventional platform intermediators, an open source blockchain with PoW as well as Interlaces' permissioned blockchain approach, are compared. The following confrontation is based on deductive criteria, which illustrates the inherent characteristics of a commons-based sharing economy.

## 1 INTRODUCTION

The sharing economy has gained increasing social acceptance, driven by socio-ecological, economic and technological changes (Botsman 2013, online). It has the potential for more sustainable business using digital transformation (Baier et al. 2016, 23-29) by shifting emphasis on open networks (Benkler 2006, 4-5), trust and more efficient resource usage (Siefkes 2016, 50, Baier et al. 2016, 35-38).

However, the practice of the sharing economy by conventional platform providers shows an intensification of capitalism. In a grey area between market and state regulation (Sundararajan 2016, 3, 26-27) platform providers enrich themselves (Reillier, Reillier 2017, 2), without assuming responsibility for safety or local regulations (Slee 2015, 49-53; Martin 2016, 153).

To ensure that the potential for more sustainable business is not lost in the platform economy, a commons suitable implementation must be found (Klapper, Martin, Upham 2017, 1395). For this purpose, this paper examines a blockchain-based open source Peer-to-Peer (P2P) mediation, which enables direct interaction of peers on the basis of democracy and self-preservation (Nakamoto 2008, 1-3). The corporation Sardex in cooperation with the EU-research project Interlace offers a complementary market based on a blockchain application. Due to them taking responsibility for social and local issues, they are used as case study in this paper.

The main contributions of this paper are to investigate the implementation compatibility of a commons-based sharing economy through a blockchain-based open source mediation and to identify deductive criteria, which illustrate the inherent characteristics of such a commons-based sharing model. Therefore, the network economic context of the sharing economy with its sociological and technological drivers is analysed and the different implementation options through conventional platform intermediators, a blockchain with PoW and Interlaces´ permissioned blockchain approach are compared in regard to their commons-compatibility.

## 2 CONCEPTUAL BACKGROUND

Extensive debates in different contexts lead to various research focuses and interpretations of the sharing economy. This contribution focusses on high impact sharing economy concepts and divides them into a commons and a platform-based sharing economy.

### 2.1 Related Work

The main bulk of sharing economy research is focused on network economic concepts and business models, while other researchers predict fundamental changes in economic paradigms.

Sundararajan (2016), Evan and Schmalensee (2016) as well as Reillier and Reillier (2017) analyze the usage of network effects within the sharing economy and which business models can be derived from it. Tirole (2017) delves towards technical issues of two-sided platforms, concerning the transformation of business, work and regulation.

Botsman (2013) recognizes technological, social, ecological and economic drivers of the sharing economy and that its success is based on convenience, transparency and participation.

Benkler (2006) predicts a revolution of our economic paradigm through cooperation and allocation of information goods to marginal costs, close to zero. Rifkin (2014) expands upon this concept by digital transformation and the inclusion of the communication, energy and logistics infrastructure, resulting in disappearing marginal costs, also for tangible goods.

Driven by cultural changes at the beginning, Bardhi and Eckhard (2012) investigate the shift from ownership to access and ascertain, that sharing economy is motivated by cost-effectiveness and convenience (2015). Siefkes (2016) criticizes that, while social welfare improvements can be realized through responsible usage of technology in the sense of Ostroms commons, negative effects on the labor market and contradictions of fundamental cost-cuts by investment-intensive goods will arise.

Sharing economy as a pure intensification of the capitalistic system is represented by Slee (2015). Non abidance of laws, profit maximization of individuals and exploitation of communities in combination with exponential growth are challenging social states.

This paper builds on the introduced work as well as the academic work "Eine Commons-gerechte Umsetzung der Sharing Economy" (Unterberger 2019) to investigate a commons-compatible implementation of the sharing economy.

### 2.2 Terminology - Sharing and Network Economy

The sharing economy is positioned between market and state planning by using the decentral internet structure. It enables the direct exchange (Rifkin 2014, 342; Slee 2015, 9) of unused digital and physical resources (Botsman 2013, online) between members of digital communities on a high scale (Sundararajan 2016, 38) (Table 1, S1.1, S3.4). Commons of the sharing economy are characterized by non-exclusion (Moglen 1999, 21-22; Benkler 2006, 61-63) and decreasing rivalry in consumption (Merten, Meretz 2005, 305-309). Their purpose is to sustain members with useful goods. Revenues and costs from usage and contribution are generated and allocated democratically and self-governing within the community (Ostrom 1999, 116-118; Siefkes 2016, 51-52) (Table 1, S2.3, S3.1 and S3.4). In comparison, the platform economy is market based and capitalistic, organized via crowd-based networks (Parker, Van Alstyne, Choudary 2016, 15). Thus it enters a grey area between private and business exchange (Sundararajan 2016, 26-27).

Due to the core of the sharing economy being its decentralized character, it is dominated by the network economy (Evans, Schmalensee 2016, 21). Two-sided markets, where sellers and customers can interact with each other, aim to use positive, indirect network effects, which attract new members on the supply as well as on the demand side and avoid negative externalities (Parker, Van Alstyne, Choudary 2016, 29-31, Tirole 2017, 379-387), resulting in a positive feedback loop. Such demand based economies of scale are responsible for the disproportionate growth of successful sharing models (Shapiro, Varian 1999, 174) (Table 1, T1).

## 3 DRIVERS OF A COMMONS-BASED SHARING ECONOMY

Social change and technical innovations depend on each other. In order to keep the focus on these two drivers of the sharing economy, ecological and economical aspects are not explicitly discussed. The sharing economy was developed as a niche in different areas of life, evolved to bring regimes into question with its social values, culture and economic paradigm (Martin 2016, 149-150, 158; Baier et al. 2016), as well as close to zero marginal costs, open source software and digital transformation (Benkler 2006, 52; Rifkin 2014, 107).

Preprint## 3.1 Sociological Drivers

The commons-based sharing economy empowers peers worldwide to create, share and develop together. This exchange is based on trust and social capital (Cherry, Pidgeon 2018, 939-940), which is primarily generated by decentralized reputation systems and photographs (Ert, Fleischer, Magen 2016, 63) (Table 1, S4.1, S4.2 and S4.4).

Moreover, a cultural shift can be recognized from ownership, which loses its status function, to access, which enable flexibility for a rapidly changing, dematerialized society (Baradhi, Eckhardt 2012, 883). This is shown by collaborative consumption, where time, resources or skills are shared and exchanged directly between peers (Rifktin 2014, 329-330) (Table 1, S4.3 and T4.3).

Even the economic paradigm is challenged by the emergence of the democratic and self-organized commons-based sharing economy. Commons realize their maximum value due to non-exclusion and self-management (Rose 1986, 774, 779-781). According to Ostrom (1999, 116-118) this requires certain rules which can be adapted to global peer production. Key roles are the consideration of local circumstances, direct democracy, transparent structures and governmental recognition (Benkler 2006, 3, 275) (Table 1, S1, T2.3, S3.2, S3.3 and S4.2).

However, within the platform economy, contributions are not shared, but made available by micro-entrepreneurs for an untaxed fee (Martin 2016, 153) with reputation systems being criticized to be inadequate in replacing safety or hygiene regulations (Slee 2015, 117-130). Moreover, platform economists use access for cost efficiency as well as convenience and apply government regulations to their own benefit (Bardhi, Eckardt 2015, online), while owners can arbitrarily decide which peers they grant access (Slee 2015, 49). This is highlighted by Hardin (1968), who in comparison to Ostrom (1968, 1244-1248) assumes the failure of commons in the long term, since humans primarily pursue their individual benefits. Additionally, users have no say or control but have to bear the risk of breaches of agreements or laws (Slee 2015, 52-53; Klapper, Martin, Upham 2017, 1395).

## 3.2 Technological Drivers

Information goods are the basis of the sharing economy and can already be produced and distributed at almost zero marginal costs, which describes optimal productivity (Rifkin 2014, 12-14, 18), devolving power from resource scarcity and making human communication capacity the key resource (Benker 2006, 52) (Table 1, T1.1 and T4.4).

Table 1: Categorization of commons-based sharing economy variables (Interpretation of the introduced sociological and technological drivers)

|  | S Sociological | T Technological |
|---|---|---|
| 1 Economic Focus | S1.1 Positioning between market and state – Combination of value creation and sharing | T1.1 Two-sided networks – Demand-based economies of scale, Positive externalities |
|  | S1.2 Self-management | T1.2 P2P networks |
|  | S1.3 Democracy | T1.3 Distributed Platform |
| 2 Users and Resources | S2.1 Distinction of users and resources | T2.1 Open source license |
|  | S2.2 Control of users and resources – Simple conflict resolution mechanisms | T2.2 Traceability and transparency – Reputation systems and safety standards |
|  | S2.3 Economic commons – Goods and services, reducing rivalry | T2.3 Digital commons – Digitalization of goods and services, marginal costs near zero |
| 3 Institutional Rules | S3.1 Allocation and acquisition | T3.1 Transaction flow |
|  | S3.2 Coherence with local conditions | T3.2 Differentiation of local and global networks |
|  | S3.3 Recognition of the state | T3.3 Legal compliance |
|  | S3.4 Supply of stakeholders – Unused resources | T3.4 Availability |
| 4 Social Values | S4.1 Social capital – Sociological trust | T4.1 Social welfare – Technological trust |
|  | S4.2 Access – Non-exclusion, sharing | T4.2 Open source software |
|  | S4.3 Collaborative consumption | T4.3 Direct P2P exchange |
|  | S4.4 Social interaction | T4.4 Communication capacity as key resource |

Open source software for example can be developed in modules from peers (Benkler 2006, 64-66), with GPL licensing e.g. ensuring that everyone has the freedom to use and develop software and that the results are subject to the same requirements (Stallmann 2015, 3). (Table 1, T2.1, T4.1 and T4.2).

The technological drivers of the sharing economy are based on the transformation of these advantages from the digital to the physical world. The combined use of the internet, apps, artificial intelligence, additive manufacturing and blockchains enables P2P exchange in real time at low transaction costs. The Internet of Things (IoT) can coordinate the communication, energy and logistics infrastructure thus also reducing marginal costs of physical goods drastically for them to become freely accessible (Rifkin 2014, 30, 36, 105-107). A free, social and highly efficient commons-based sharing economy, can replace capitalism on a large scale by turning technical progress to social progress (Rifkin 2014, 22-24) (Table 1, T2.3 and T3.4)

Nevertheless, the increase in productivity via automation not only leads to cheaper products, but also a high level of unemployment. Capitalism is largely maintained by monopolies and the marginal cost theory cannot be applied to physical goods with high fixed costs (Siefkes 2016, 40, 45-50). On the one hand, open source software and content aims to ensure freedom and social welfare. On the other hand, property rights of developers and artists are violated by illegal pirated copies (Gates 1976, online). Also, since platform economy has its focus on profit maximization (Slee 2015, 184-186), it might lead to a surveillance capitalism (Zuboff 2015, 75-76, 85).

# 4 BLOCKCHAIN-BASED MEDIATION

In addition to apps, social networks and the IoT, blockchains are technological innovations with high impact, as they can help to realize a commons-based sharing economy without the abuse of platform providers. This is because the blockchain, with its inherent P2P-structure and democratic behaviour, seems to be in line with the sociological and technological drivers discussed in chapter 3.

## 4.1 Blockchain – Definition

A blockchain is a decentralized database of single data blocks that contain transactions. Each block consists of the hash of the previous block, unconsolidated transactions and a number used once. The resulting chain displays transactional historical correctness, which can be verified through the cryptographically secure connecting hashes (Nakamoto 2008, 3-4; Hughes et al. 2018, 64).

In order to ensure open access and control, the decentralized structure must meet the requirements of consensus finding without a central instance and self-preservation (Swan 2015, 1). PoW enables this through the solution of a mathematical problem which correctness is verified by a majority. This approval is reflected by the integration of the solved data block and processing of the next block. If several blocks are sent for verification, the block accepted by the majority is preferred (Nakamoto 2008, 1, 3, Nofer, Gomber, Hinz 2017, 184), resulting in a voting process similar to direct democracy. This complex processing requires computing power, which is provided by so called miners, which are usually compensated for their service in e.g. crypto currencies (Swan 2015, 16).

## 4.2 Pro and Cons of a Blockchain

The decisive advantage of a blockchain is its decentralization and digitalization, which enables peers to exchange transactions directly (Husain, Roep, Franklin 2019, 6). Therefore, trust in intermediaries, such as platform providers becomes obsolete (Swan 2015, 17). Both, users and miners in a blockchain can act anonymously. However, studies illustrate that network analysis can identify user groups (Sixt 2017, 32-33). Maximum transparency is provided by the historical database, which can be monitored and has to be stored by all peers (Koch, Pieters 2017, 2). It is traceable and immutable due to PoW (Nakamoto 2008, 3), as long as not more than 50% of peers agree on adding a false block. The distributed data store and an increasing number of network nodes lead to scaling problems in the long term (Barber et al. 2012, 410).

Therefore, the blockchain technology is suitable to help a commons-based sharing economy to break through, as central platform providers are no longer needed, due to decisions and maintenance of the platform being directly managed by the peers at near zero marginal cost. Trust is replaced by a mathematical algorithm and anonymity makes human interaction obsolete. However, legal, energy and scaling problems must be taken into account when using blockchains as a platform for sharing economy.

## 5. EVALUATING THE COMMONS-COMPATIBILITY OF A BLOCKCHAIN APPROACH

In order to investigate, if a commons-compatible implementation of the sharing economy is possible using blockchain technology, a qualitative case study is performed. Therefore, Sardex in cooperation with Interlace and its blockchain application are empirical evaluated.

### 5.1 Methodology – Qualitative Case Study

A structured content analysis according to Mayring (2015) is used as qualitative evaluation method due to the novelty of the issue and therefore needed open, explorative access and consideration of the social context.

Therefore, the company Sardex S.p.A. in cooperation with the Horizon 2020, EU-research project Interlace (No. 794494) and their blockchain application are selected for a case study. Qualitative data material, such as expert interviews with P. Dini (2019) and E. Hirsch (2019) as well as the whole Sardex and Interlace consortium are analysed. The two academic experts have objective, specialized key-knowledge, due to their research, consulting and control functions within Sardex and Interlace.

The case is evaluated based on criteria for a commons-based implementation of the sharing economy. These criteria, which can be seen in Table 1, are derived from the drivers and the definitions of a commons-based sharing economy and expanded trough the evaluation of the Sardex and Interlace case study. The variables are structured via deduction of terms regarding intention and extension (Tatievskaya 2005, 53-54) and illustrated in a multidimensional category system. Due to strong relations between sociological and technological criteria, they are analysed in reference to each other and further form four main rubrics. The category "Economic Focus" deals with its economic position, management and maintenance issues. "Users and Resources" includes the necessary characteristics of participants, goods and services regarding their commons-compatibility. How these users get in touch and deal with each other as well as how these resources are allocated and acquisitioned is defined within "Institutional rules". Issues regarding interaction and allocation beyond institutional rules with the focus on trust and sharing in the sense of commons are discussed within "Social values". The rehashed data is analysed, assigned to the appropriate variables and evaluated with a qualitative content analysis.

### 5.2 Sharing Economy by Sardex and Interlace

This section evaluates the realization of the sharing economy by Sardex in cooperation with Interlace and its blockchain application empirical in reference to the deductive commons-based criteria.

Under an economic focus, Sardex offers a complementary market based on mutual credits, trust, goods and services with a virtual network as a key resource. Through the Interlace blockchain, interest free loans become traceable and scale to other regions with almost zero transaction costs (Dini et al. D2.2 2018, 13-15; Dini, Hirsch 2018, D3.2, 12). Each peer can submit or reject each transaction, copy the entire chain and disconnect from the system at any time (Hirsch 2019, 8-11). Sardex ensures distributed control, but does not implement a completely open blockchain to pursue social values and the monetary interests of investors (Dini 2019, 2, 6-7) (Table 1, T3.2).

Users and resources are controlled central by Sardex. Registered companies are distinguished based on their communication and economic as well as financial interaction possibilities (Dini et al. D3.1 2018, 9-12).

Institutional rules define the allocation and acquisition of mutual loans within Sardex. They are provided on the basis of real economic potential (Dini 2019, 9) and implemented via smart contracts. (Dini et al. D3.1 2018, 7-8). A balance between local and global requirements is technologically supported through distinction of a community, application and infrastructure Layer. Legally prescribed taxes and rules are paid directly and observed (Dini et al. D2.3 2018, 8). (Table 1, S2.1, S2.3, S3.3, T3.3 and T3,5).

Sardex and Interlace focus on social values, which are based on ethical considerations and implemented through judgmental technology. The foundation of Sardex is sociologically and technologically trust. Furthermore, it prioritizes solidarity, local culture and mutuality, which are implemented through controlled technology by Interlace (Dini et al. D2.3 2018, 10-12; Dini 2019, 11-12) which is open and free of charge for copies, modifications, extensions and publications to everyone (Dini D1.1 2017, 4-5) (Table1, S4.1 and T4.1).

# 6 FINDINGS – BLOCKCHAIN-BASED MEDIATION

This section compares Sardex in cooperation with Interlace in regard to the theoretically derived criteria, crucial for a commons-based sharing economy, with the possibilities of a permissionless blockchain as well as a typical platform economic system to get the most compatible one. To which extend a blockchain-based open source mediation enables a commons-compatible practice, is answered by linking the introduced theory to empiricism.

Economic focus: A permissionless blockchain, Sardex with Interlace and traditional sharing platforms, each generate value by providing peers the ability to transact directly with each other in almost real time. While a permissionless blockchain offers this completely distributed, Sardex provides this via a complementary market with mutual credits and platform economy in a capitalistic grey area between commerce and private business. While peers manage and maintain the permissionless blockchain by themselves, central intermediaries charge high commissions. The centrally managed Sardex, e.g. collects an annual membership fee for maintaining the technological infrastructure (Dini, Hirsch 2018, D3.2, 41). An open blockchain with PoW is governed democratically, while neither platform providers nor Interlace can make majority decisions.

Users and Resources: The inclusivity of a permissionless blockchain also comes with problems, since it is unable to exclude peers and traded content is uncontrollable. In contrast, Sardex gives its participants access to interest-free credit while supporting social interaction. Platform sharing models usually give strangers access without social interaction or respect to the needs of the commune.

Institutional Rules: Control in blockchains is completely distributed and trust is replaced by algorithms, whereas trade in the platform economy is based on P2P rating systems. While Sardex complies with all laws as well as tax regulations and is recognized by the state and international authorities, a permissionless blockchain only implements regulations that are passed by the majority of peers. In contrast, platform intermediators try to shift the responsibility for law-abiding or transparent transactions to its peers.

Social Values: Blockchains with their own currencies in tokens tend to be subject to speculation. In contrast, Sardex with Interlace offers trade on basis of actual economic potential by maximal transparency. Nakamoto (2008), with his publication of the blockchain through Bitcoin, makes trust in intermediaries obsolete and revolutionizes P2P exchange, while Sardex is based on trust and takes supra-regional rules and local culture into account. In contrast, traditional sharing platforms have no regard for the needs of the commune.

The compatibility of a commons-based implementation of the sharing economy through blockchain-based open source mediation varies between different criteria. While it meets the variables concerning the economic focus best and is compatible with most of the institutional rules, social values suffer under the direct democracy and inclusivity, which leads to issues regarding self-serving users, who exploit the system for trading e.g. illegal resources. This shows that a blockchain based open source mediation on itself is incompatible with a commons-based implementation of the sharing economy.

# 7 RECOMMENDED COMMONS-COMPATIBLE IMPLEMENTATION OF THE SHARING ECONOMY

Based on the findings in chapter 6 and under the prevailing conditions, that individuals act opportunistic within a capitalistic system, a combination of a permissionless blockchain and a publicly voted central system is recommended. This results in a suitable technology that ensures traceability, distributed control and takes local and social considerations into account.

While consensus building through PoW can be used to participate in a direct democratic way regarding commons investments and mediation, an elected group of stakeholders can take responsibility for law-abidance and commons compatibility, like representative democracy. Thus, decentralised majority voting is used to integrate peers into the administration and to make them aware of their responsibility towards society. To enable mutual allocation and acquisition of tangible commons, such as 3D printers or solar panels, they can be financed by investing a percentage of the transaction volume in order to aid most members. Further, such mutual and effective collaboration enables investments in state-of-the-art technology. This aims to attract more users and lead to positive, indirect network effects. The group of representatives are also elected by the majority and financed by fees, which are automatically paid by transactions. The amount of fee is determined by the members democratically and can be checked in the historical database.

In summary, a commons-compatible implementation of the sharing economy is possible by combining a permissionless blockchain-based open source software with a publicly voted central system. While PoW enables majority voting, distributed control and transparency, a group of local stakeholders are able to take responsibility of laws and cultures in the sense of commons.

# 8   CONCLUSION

This paper deals with the drivers of the network economical commons-based sharing economy and investigates the various compatibilities by comparing different implementations through conventional platform intermediators, an open source blockchain with PoW and the permissioned blockchain of Sardex in cooperation with Interlace in regard to commons-characteristic criteria. Based on that confrontation, the compatibility of a commons-based implementation of the sharing economy through a blockchain-based open source mediation is explored.

To summarize, network effects enable rapid development of both, commons-based and platform economic sharing economy. Blockchain technology, with its decentralized, historical database and consensual finding with PoW seems to have potential for a commons-compatible implementation of the sharing economy. However, the single case study of Sardex in cooperation with Interlace and their blockchain application illustrates, that in practice the distributed, permissionless blockchain only does this to a limited extend. This neutral technology realizes democracy and independence from intermediators. Therefore, it is subject to the will of the majority, who prefer monetary benefit maximization under anonymity. In comparison, a permissioned blockchain, used for distributed control and transparency, supports a more commons-compatible implementation, as long as the central authority takes social responsibility. Most contradictions result from the implementation of a commons-based sharing economy through platform economy, but even the adapted blockchain application cannot meet all commons criteria, as the evaluation of the paper shows. Therefore, a combination of a permissionless blockchain-based open source mediation with a publicly voted central system is recommended for a commons-compatible implementation of a sharing economy based on transparency, democracy and taking supra regional as well as local social needs into account.

The conclusion, that direct democratic technology fulfils most of the economic and institutional requirements, but does not lead to a commons-based trade, has consequences. It results that a commons-compatible implementation is not advantageous for majority of peers, or that they are unable to estimate the long-term effects of current decisions, thereby reaching limits of direct democracy. Both, this recognition as well as the prognosis that the platform economy will continue within our capitalistic system, makes further research obvious. This includes the analysis of the effects of progressive automation on the labour market as well as the consequences of permanent data collection and use within capitalism.

# ACKNOWLEDGEMENT

We thank Hermann Rauchenschwandtner, FH-Prof. Dr. Dr., Salzburg University of Applied Sciences, who provided insight and expertise that greatly assisted the research.

# REFERENCES

Baier, A., Hansing, T., Müller, C., Werner, K. (Hg.) (2016): Die Welt reparieren. Open Source und Selbermachen als postkapitalistische Praxis. Bielefeld: Transcript

Barber, S., Boyen, X., Shi, E., Uzun, E. (2012): Bitter to better. How to make bitcoin a better currency. In: International Conference on Financial cryptography and Data Security. Heidelberg: Springer, 486-504

Bardhi, F., Eckhardt, G. M. (2012): Access-Based Consumption. The case of Car Sharing. In: Journal of Consumer Research. 39 (2012), 881-898

Bardhi, F., Eckhardt, G. M. (2015): The Sharing Economy Isn´t About Sharing at All. In: Harvard Business Review. https://hbr.org/2015/01/the-sharing-economy-isnt-about-sharing-at-all. Access at 28.10.2018

Benkler, Y. (2006): The Wealth of Networks. How Social Production Transforms Markets and Freedom. New Haven, London: Yale University Press

Botsman, R. (2013): The Sharing Economy Lacks A Shared Definition. In: Fastcompany. https://www.fastcompany.com/3022028/the-sharing-economy-lacks-a-shared-definition. Access at 16.08.2018

Botsman, R., Rogers, R. (2010): What´s Mine Is Yours. The Rise of Collaborative Consumption. New York: HarperCollins

Cherry, C. E., Pidgeon, N. F. (2018): Is sharing the solution? Exploring public acceptability of the sharing economy. In: Journal of Cleaner Production. 195 (2018), 939-948

Dini, P. (2019): A Commons-Compatible Implementation of the Sharing Economy. Expert Interview

Dini, P. D1.1 (2017): Project Website. In: WP1. Coordination, Management and Dissemination.

Dini, P., Hirsch, E. D3.2 (2018): Final Demonstrator Implementation. In: WP3. Iterative Demonstrator Implementation. (Access Deliverables)

Dini, P., Hirsch, E., Börger, E., Mulas, M. L. D3.1 (2018): First Demonstrator Implementation. In: WP3. Iterative Demonstrator Implementation. (Access Deliverables)

Dini, P., Hirsch, E., Littera, G., Carboni, L. D2.2 (2018): Iterative Architecture Refinement. In: WP2. Iterative Architecture Requirements and Definition. (Access Deliverables)

Dini, P., Hirsch, E., Littera, G., Carboni, L. D2.3 (2018): Final Architecture. In: WP2. Iterative Architecture Requirements and Definition. (Access Deliverables) Access Deliverables at 17.02.2019: https://www.interlaceproject.eu/#deliverables.

Ert, E., Fleischer, A., Magen, N. (2016): Trust and reputation in the sharing economy: The role of personal photos in Airbnb. In: Tourism Management. 55 (2016), 62-73

Evans, D. S., Schmalensee, R. (2016): Matchmakers. The New Economics of Multisided Platforms. Boston: Harvard Business Review Press

Gates, B. (1976): An Open Letter to Hobbiests. In: Genius. https://genius.com/Bill-gates-an-open-letter-to-hobbyists-annotated. Access at 04.01.2019

Hardin, G. (1968): The Tragedy of the Commons. In: Science. 162 (3859), 1243-1248

Hirsch, E. (2019): A Commons-Compatible Implementation of the Sharing Economy. Expert Interview

Hughes, E., Graham, L., Rowley, L., Lowe, R. (2018): Unlocking Blockchain: Embracing New Technologies to drive Efficiency and Empower the Citizen. In: The JBB. 1(2), 63-73

Husain, S. O., Roep, D., Franklin, A. (2019): Prefigurative Post-Politics as Strategy: The Case of Government-Led Blockchain Projects. In: The JBBA. 3(1), 1-11

Klapper, R., Martin, C. J., Upham, P. (2017): Democratising platform governance in the sharing economy: An analytical framework and initial empirical insights. In: Journal of Cleaner Production. 166 (2017), 1395-1406

Koch, C., Pieters, G. C. (2017): Blockchain Technology Disrupting Traditional Records Systems. In: Financial Insights. 6 (2), 1-4

Martin, C. J. (2016): The sharing economy. A pathway to sustainability or nightmarish form of neoliberal capitalism? In: Ecological Economics. 121 (2016), 149-159

Mayring, P. (2015): Qualitative Inhaltsanalyse. Grundlagen und Techniken. 12th, revised edition. Weinheim, Basel: Beltz

Merten, S., Meretz, S. (2005): Freie Software und Freie Gesellschaft. In: Bärwolff, M., Gehring, R., Lutterbeck, B. (Hg.): Open Source Jahrbuch 2005. Zwischen Softwareentwicklung und Gesellschaftsmodell. Berlin: Lehmanns Media, 293-309

Moglen, E. (1999): Anarchism Triumphant. Free Software and the Death of Copyright. In: First Monday. 4 (8). http://journals.uic.edu/ojs/index.php/fm/article/view/684. Access at 28.10.2018.

Nakamoto, S. (2008): A peer-to-peer electronic cash system. https://bitcoin.org/bitcoin.pdf. Access at 07.01.2019

Nofer, M., Gomber, P., Hinz, O., Schiereck, D. (2017): Blockchain. In: Bus & Inf Syst Eng. 59 (3), 183-187

Ostrom, E. (1999): Die Verfassung der Allmende. Jenseits von Staat und Markt. Tübigen: Mohr Siebeck

Parker, G. G., Van Alstyne, M. W., Choudary, S. P. (2016): Platform Revolution. How Networked Markets are Transforming the Economy – and how to Make Them Work for You. London, New York: W. W. Norton & Company

Reillier, L. C., Reillier, B. (2017): How to Unlock the Power of Communities and Networks to Grow Your Business. London, New York: Routledge

Rifkin, J. (2014): Die Null Grenzkosten Gesellschaft. Das Internet der Dinge, kollaboratives Gemeingut und der Rückzug des Kapitalismus. Frankfurt, New York: Campus Verlag

Rose, C. M. (1986): The Comedy of the Commons: Commerce, Custom, and Inherently Public Property. In: The University of Chicago Law Review. 53 (3), 711-781

Shapiro, C., Varian, H. R. (1999): Information Rules. A Strategic guide to the Network Economy. Boston, Massachusetts: Harvard Business School Press

Siefkes, C. (2016): Freie Software und Commons. Digitale Ausnahme oder Beginn einer postkapitalistischen Produktionsweise? In: Medienwissenschaften und Kapitalismuskritik. 16 (2), 37-54

Sixt, E. (2017): Bitcoins und andere dezentrale Transaktionssysteme. Blockchains als Basis einer Kryptoökonomie. Wien: Springer Gabler

Slee, T. (2016): Deins ist Meins. Die unbequeme Wahrheit der Sharing Economy. München: Antje Kunstmann

Stallman, M. (2015): Free Software, Free Society. 3th edition. Boston: Free Software Foundation

Sundararajan, A. (2016): The Sharing Economy. The End of Employment and the Rise of Crowd-Based Capitalism. Cambridge, London: The MIT Press

Swan, M. (2015): Blockchain. Blueprint for a New Economy. Sebastopol, CA: O´Reilly Media

Tatievskaya, E. (2005): Der Begriff der logischen Form in der analytischen Philosophie. Russell in Auseinandersetzung mit Frege, Meinong und Wittgenstein. Frankfurt, Paris, Ebikon, Lancaster, New Brunswick: Ontos

Tirole, J. (2017): Economics for the Common Good. New Jersey: Princeton University Press

Unterberger, P. (2019): Eine Commons-gerechte Umsetzung der Sharing Economy.

Zuboff, S. (2015): Big other: surveillance capitalism and the prospects of an information civilization. In: Journal of Information Technology. 30 (1), 75-89